\documentclass[conference]{IEEEtran}

\usepackage{amssymb}

\usepackage{floatflt}

\usepackage{cite}

\usepackage{graphicx}
\usepackage{verbatim}

\usepackage{psfrag}

\usepackage{subfigure}

\usepackage{url}

\usepackage{stfloats}

\usepackage{amsmath}
\usepackage{algorithmic}
\usepackage{amsthm}

\IEEEoverridecommandlockouts
\IEEEpubid{\makebox[\columnwidth]{978-1-4799-5863-4/14/{\$}31.00\copyright2014 IEEE \hfill} 
\hspace{\columnsep}\makebox[\columnwidth]{ }}

\begin{document}

\title{Effective Capacity of Cognitive Radio Links: Accessing Primary Feedback Erroneously}

\author{\authorblockN{M. Majid Butt\authorrefmark{1}, Ahmed H. Anwar\authorrefmark{1}, Amr Mohamed\authorrefmark{1} and Tamer ElBatt\authorrefmark{2}}\\
\authorblockA{\authorrefmark{1}Department of Computer Science and Engineering, Qatar University, Qatar\\
E-mail: \{majid.butt, a.anwar, amrm\}@qu.edu.qa}
\authorblockA{\authorrefmark{2}Nile University, Egypt, E-mail: telbatt@ieee.org
}}

\maketitle

\begin{abstract}
We study the performance of a cognitive system modeled by one secondary and one primary link and operating under statistical quality of service (QoS) delay constraints. We analyze the effective capacity (EC) to quantify the secondary user (SU) performance under delay constraints. The SU intends to maximize the benefit of the feedback messages on the primary link to reduce SU interference for primary user (PU) and makes opportunistic use of the channel to transmit his packets. We assume that SU has erroneous access to feedback information of PU. We propose a three power level scheme and study the tradeoff between degradation in EC of SU and reliability of PU defined as the success rate of the transmitted packets. Our analysis shows that increase in error in feedback access causes more interference to PU and packet success rate decreases correspondingly.
\end{abstract}
\vspace{-0.2cm}
\section{Introduction}
\label{sect:Int}
Scarcity of bandwidth has resulted in increased research interest in the area of cognitive radio (CR) networks. 
Cognitive radios aim to provide efficient spectrum utilization as the SUs access the (under used) spectrum assigned to the licensed PUs without compromising QoS for the PUs.

Information \emph{capacity} is defined as the tightest upper bound on the amount of information that can be reliably transmitted over a communications channel. In real-time applications, guaranteeing a QoS constraint is essential. Wu {\it et al.} introduced a different notion of capacity under QoS constraint in \cite{wu2003effective}, namely, \emph{effective capacity} for a wireless channel. It is defined as the maximum constant arrival rate that can be supported by a given channel service process while satisfying a statistical QoS requirement specified by the QoS exponent. This concept was introduced for link layer modeling in order to capture QoS requirements, e.g., delay, such that performance analysis in wireless systems can be performed without going into complex queuing analysis. The notion of EC measures the system throughput when it is conditioned by a QoS constraint. EC is a dual concept in wireless communication arena to the concept of \emph{effective bandwidth} which was originally introduced for wired networks \cite{chang1995effective}.
The authors in \cite{musavian2010effective} investigate the interference and delay constrained CR relay channels while multi-user case is investigated in \cite{shakkottai2008effective} and a formulation of the EC with QoS constraints is proposed. The authors provide two scheduling policies and show that both algorithms yield the same long-term throughput in the absence of QoS constraint but result in widely different throughput if they are imposed by QoS constraints. However, none of these studies consider the effect of feedback in terms of EC notion.

EC is a system performance metric to capture CR fundamental tradeoff between SU throughput and the corresponding outage caused to PU in a delay sensitive system. EC of SU exploiting PU feedback was investigated in \cite{mypaper}. The work assumes a perfect feedback channel between primary transmitter (Tx) and receiver (Rx). The feedback channel contains the information about successful/unsuccessful reception of the packet.
If a packet is received at the receiver, no explicit acknowledgement (ACK) message is fed back, otherwise a negative acknowledgement (NACK) is transmitted. The SU has access to this feedback channel and exploits this information to improve EC.
The aims of this study are twofold:
\begin{itemize}
  \item {We propose a scheme which outperforms the scheme presented in \cite{mypaper} in terms of packet success rate of the PU due to less interference caused by the SU. Minimum (or no) interference to PU is the fundamental principle behind the idea of CR concept. In contrast to solely SU EC analysis presented in \cite{mypaper}, we analyze both SU EC and PU packet success rate to determine the tradeoff between primary and secondary network performance.}
  \item {We extend the analysis to more practical case where SU is able to access PU feedback information with probability $1-\epsilon$ where $\epsilon$ is probability that PU receives the feedback but SU cannot access it. This phenomenon is similar to well known hidden terminal problem. Our contribution is to quantify the effects of erroneous access to feedback information on both EC of SU and success rate of PU.}
\end{itemize}
\IEEEpubidadjcol

The rest of the paper is organized as follows. The system model and underlying assumptions are presented in Section \ref{sysmod}. In Section \ref{sect:Schemes}, the EC problem for our proposed scheme is formulated and analyzed. Afterwards, numerical results and discussion are presented in Section \ref{sect:results}. Finally, we conclude with the summary of the main contributions in Section \ref{sect:conclusions}.

\section{System Model}\label{sysmod}
We consider a time slotted system. The primary network is abstracted by a primary link (i.e, our analysis is valid for any number of PUs). The primary Tx accesses the channel whenever it has a packet to send in its queue. On the other hand, a single SU attempts to access the medium with a certain policy based on the spectrum sensing outcome. The SU is assumed to have a packet to send at the beginning of each time slot (fully backlogged). Data is transmitted in frames of duration $T$ seconds, where each frame fits exactly in a single time slot. We assume that the first $N$ seconds of the frame duration $T$ are used by the SU to sense the licensed spectrum.

The discrete time secondary link input-output relations for idle and busy channels in the $i^{th}$ symbol duration, respectively, are given by
\begin{eqnarray}
y(i)&=&h(i)x(i)+n(i) \qquad  i=1,2, \cdots \\
y(i)&=&h(i)x(i)+s_p(i)+n(i) \quad  i=1,2,\cdots, \label{relation2}
\end{eqnarray}
where $x(i) $ and $ y(i)$ represent the complex-valued channel input and output, respectively. $h(i)$ denotes the channel coefficient between the cognitive transmitter and receiver, $s_p(i)$ is the interference coefficient from the PU to the SU and $n(i)\sim \Psi(0,\sigma_n^2)$ is the Gaussian noise with zero mean and variance $\sigma_n^2$. The channel bandwidth is denoted by $B$. The channel input is subject to the average energy constraint: $\mathbb{E}{\lbrace|x(i)|^2}\rbrace\leq P_j/B$ with $j\in\{0,1,2\}$ and $P_j$ denoting SU power level depending on the sensing outcome and PU activity as explained later.
The fading coefficients are assumed to have arbitrary marginal distributions with finite variances such that
\begin{equation}
 \mathbb{E}{\lbrace|h(i)|^2}\rbrace = \mathbb{E}\lbrace z(i) \rbrace =  \sigma ^{2}< \infty
 \end{equation}
where $|h(i)|^2 = z(i)$.
Finally, we consider a block-fading channel model and assume that the channel stays constant for a block of duration $T$ seconds and vary independently from one block to another.

We adopt the cognitive sensing and outcome framework in \cite{musavian2010effective, mypaper}. We assume that PU occupies the wireless channel with a fixed prior probability $\rho$. The channel sensing can be formulated as a hypothesis testing problem between the additive white Gaussian noise $n(i)$ and the primary signal $s_p(i)$. As there are $NB$ complex symbols in a duration of $N$ seconds, this can be expressed mathematically by
\begin{eqnarray}
H_0 &:& y(i)=n(i), \qquad i=1,...,NB;\\
H_1 &:& y(i)=s_p(i)+n(i), \quad i=1,...,NB.
\end{eqnarray}
The probabilities of false alarm (FA) $P_f$ and mis-detection (MD) $P_d$ are computed by
\begin{eqnarray}\label{prob_f}
P_f&=&Pr(Y> \lambda | H_0)=1 - P\left ( \frac{NB\lambda}{\sigma ^{2} _{n} }  , NB \right)\\
P_d&=&Pr(Y> \lambda | H_1) =1 - P\left ( \frac{NB\lambda}{\sigma ^{2} _{sp}+\sigma ^{2} _{n} }  , NB \right)
\label{prob_d}
\end{eqnarray}
where $ \lambda $ is the energy detector threshold, $Y = \frac{1}{NB} \sum_{i=1}^{NB} |y(i)|^2$ and $P(x, a)$ denotes the regularized lower gamma function defined as $P(x, a)=\frac{\gamma (x,a)}{\Gamma(a)}$ with $\gamma (x,a)$ denoting the lower incomplete gamma function.

Taking sensing errors into account according to (\ref{prob_f}) and (\ref{prob_d}), the sensing process results in one of the following outcomes:
\begin{enumerate}
\item Channel busy--detected busy, denoted by (B-B)
\item Channel busy--detected idle, denoted by (MD)
\item Channel idle--detected busy, denoted by (FA)
\item Channel idle--detected idle, denoted by (I-I)
\end{enumerate}

\subsection{Preliminaries For Effective Capacity Analysis}
\label{sect:EC analysis}
This section summaries the preliminaries for EC analysis from \cite{musavian2010effective, mypaper} and introduces the framework used in analysis later.

Approximating the PU interference term on the SU, $s_p(i)$, as an additional Gaussian noise, we can express the SU instantaneous channel capacities in five scenarios as follows:
\begin{equation}
C(i)_l=B\log \big(1+{\rm SNR}_l z(i)\big), \;\;\;\; l=1,\dots 5,
\end{equation}
where $ {\rm SNR}_1= \frac {P_1}{B(\sigma_n^2+\sigma^2_{s_p})} $, $ {\rm SNR}_2= \frac {P_2}{B(\sigma_n^2+\sigma^2_{s_p})} $, $ {\rm SNR}_3= \frac {P_1}{B(\sigma_n^2)} $, $ {\rm SNR}_4= \frac {P_2}{B(\sigma_n^2)} $ and $ {\rm SNR}_5= \frac {P_0}{B(\sigma_n^2+\sigma^2_{s_p})} $.

We assume that the SU transmitter has no channel state information to set the transmitted data rate in every slot. Hence, the transmission rate of the SU may be smaller or greater than the instantaneous channel capacity $C(i)$. Following the framework in \cite{mypaper,akin2010effective}, the channel can be either ON or OFF. If the transmission rate is smaller than the instantaneous channel capacity, the channel is said to be ON; otherwise, the channel is considered in the OFF state (outage state). When the channel is OFF, reliable communication is not attained and the information has to be resent. We incorporate a  simple automatic repeat request (ARQ) mechanism to acknowledge the reception of data as explained in Section \ref{sect:Int}. Accordingly, the effective transmission rate in the OFF states is zero.

We develop a Finite state Markov chain (FSMC) model to characterize the sensing outcomes and feedback access reliability for SU. Additionally, Markov chain captures the channel state (ON or OFF) as well.
The Markov chain is fully characterized by its state transition probability matrix $\mathbf{R}_{M \times M}$ defined as:
\begin{equation}\label{StateTr}
\mathbf{R}_{M \times M} =
\begin{bmatrix}
p_{i,j}
\end{bmatrix},  1 \leq i,j \leq M,
\end{equation}
where $M$ is the number of states in the Markov chain. Along the same lines of \cite{akin2010effective}, the EC for such system model is expressed as follows\footnote{The proof can be found in \cite[Ch.7]{chang2000performance}. The details of EC computation using spectral radius of the matrix are omitted to focus on main idea and avoid repetition of discussion in \cite{mypaper,akin2010effective}.}:
\begin{equation}
EC(\theta)=\frac{\Lambda(-\theta)}{-\theta}=\max_{r_0,r_1,r_2} \frac{1}{-\theta}\log_esp(\mathbf{\Phi}(-\theta)\mathbf{R}) \label{EffectiveCapacity}
\end{equation}
where $\theta$ is QoS exponent and $\mathbf{R}$ is the state transition matrix as defined above with $sp(\mathbf{\Phi}(-\theta)\mathbf{R})$ denoting the spectral radius of the matrix $ \mathbf{\Phi}(-\theta) \mathbf{R}$ (i.e., the maximum of the absolute of all eigenvalues of the matrix). $r_j$ denotes the rate as a function of SU power $P_j$.
To reach a closed form expression for the EC, we require the eigenvalues of the matrix $\mathbf{\Phi}(-\theta) \mathbf{R}$. $ \mathbf{\Phi}(-\theta)$ is a diagonal matrix defined as $ \mathbf{\Phi}(-\theta) = diag(\phi_1(-\theta),\phi_2(-\theta),....,\phi_M(-\theta))$ whose diagonal elements are the moment generating functions of the Markov process in each of the $M$ states.
\section{Proposed Multiple Power Level Scheme}
\label{sect:Schemes}
In this section, we characterize EC of the SU and PU packet success rate for the "Double Power Level" (DPL) scheme in \cite{mypaper} and the proposed "Triple Power Level" (TPL) scheme.
First, we summarize DPL scheme:
\begin{enumerate}
         \item The PU transmits a new packet with probability $\rho$. The SU senses the medium at the start of the time slot.
\begin{itemize}
\item If the medium is sensed Busy, SU will transmit with a lower power level, $ P_1<P_2$.
\item If the medium is sensed Idle, SU will transmit with a higher power level, $ P_2 $.
\end{itemize}
         \item If the PU Tx receives a NACK from PU Rx by the end of the current time slot, the PU will retransmit the failed packet in the next time slot. After accessing PU feedback, SU will transmit with the lower power level $ P_1 $ without sensing the medium.
         \item If the PU Tx receives No feedback (implying ACK) by the end of the time slot, PU and SU perform step 1.
\end{enumerate}
The maximum number of transmissions for a single packet from PU are limited to 2 due to buffer limitations.

A fundamental principle behind the concept of CR is to provide no interference (interweave communication) or minimum interference (underlay communication). In DPL, once a NACK is received by the PU and SU has access to this information, SU should \emph{help} PU by adjusting its power level to get PU's transmission successful in the second attempt as maximum number of transmissions are limited to 2.

Based on this argument, we extend DPL scheme to TPL by introducing a third power level $P_0$ such that the second step of DPL scheme is modified as:
\begin{enumerate}
\setcounter{enumi}{1}
\item If the PU receives a NACK by the end of the current time slot, the PU transmits the failed packet in the next time slot while SU transmits with the lowest power level $P_0<P_1 $ without sensing the medium. The lower $P_0$ is, the lower is the interference caused on primary link and the higher is PU's probability of success.
\end{enumerate}
All other steps are exactly the same as in DPL scheme. One may argue that DPL can be adapted to have $P_1$ equal to (smaller) $P_0$ whenever SU senses a busy primary channel including PU's first transmission attempt and provide even less interference to PU. While this solution will provide better performance than TPL in terms of primary performance, it is more conservative approach and eliminates the opportunistic part of the scheme as compared to TPL.

To make the model more practical, we assume that PU accesses the feedback information with probability $1-\epsilon$. It is worth noting that what counts in our case is the probability of accessing a NACK message which results in a certain adaptive transmission action taken by the SU (step2). Due to our protocol, ACK is not explicit. When SU is not able to access a NACK, it mistakes this event as an ACK. Note that PU has perfect feedback channel, so erroneous feedback access case affects behavior of SU solely for that particular transmission. However, the resulting (erroneous) action taken by SU to transmit with higher power affects PU's performance in the next time slot. The result of erroneous feedback access is that the total probability of accessing a NACK by the SU is decreased by $1-\epsilon$ which in turns causes increase in interference to PU. Both DPL and TPL schemes are affected by $1-\epsilon$ factor but the severity of the effect is different as explained below.
\subsection{Effect of Erroneous Access on DPL}
For DPL scheme, accessing a NACK message leads SU to transmit with power level $ P_1 $. In the event of not being able to access NACK, SU transmits with power $P_1$ or $ P_2 $ depending on the sensing activity. PU retransmits the packet in next time slot with probability one (instead of $\rho$ as perceived by SU). If SU is able to sense PU transmission, it uses $P_1<P_2$ for transmission. Thus, in spite of failing to access feedback correctly, SU uses the correct power level and recovers well. However, if due to sensing error, SU mis-detects primary transmission corresponding to $P_d$, it will transmit with power $P_2$ which causes additional interference to PU and may cause second transmission to fail.

Thus, if the SU perfectly senses the medium, it will not lose anything due to missing the NACK message. The higher the sensing error, the higher the effect of feedback access error. In case of perfect sensing, actually system can fully recover from error in accessing the feedback and this is one of the most interesting results of this study.
\subsection{Effect on Erroneous Access on TPL}
For the TPL scheme, if the SU misses a NACK then instead of transmitting with power $P_0$, it transmits with either power $ P_1 $ or $ P_2 $ depending on the sensing result. In this case, the SU causes higher interference to PU regardless of the fact that there is a sensing error or not. Even perfect sensing as a special case cannot eliminate PU success rate degradation. Thus, the PU success rate is expected to degrade rapidly in case of TPL scheme as compared to DPL scheme. We evaluate and validate this behaviour for both DPL and TPL schemes in the numerical results further.

We characterize the PU packet success probability in the following. As packet for the PU is dropped only due to channel outage and we assume a perfect feedback channel for PU, ${\rm Pr}({\rm outage})$ equals ${\rm Pr}({\rm NACK})$ in general. Denoting SU power level by $P_{\rm sec}$, the probability of outage/NACK event for both perfect and imperfect access cases is given by
\begin{eqnarray}
\Pr({\rm NACK}_j)&=&\Pr({\rm outage}|P_{\rm sec}=P_j),\quad j=0,1,2\nonumber\\
&=&1-\Pr({\rm suc}|P_{\rm sec}=P_j)
\label{eqn:pr_nack}
\end{eqnarray}
where $\Pr({\rm suc}|P_{\rm sec}=P_j)$ and $\Pr({\rm outage}|P_{\rm sec}=P_j)$ are defined as PU packet success and PU outage probabilities, respectively when $P_{\rm sec}= P_j$. $\Pr({\rm NACK}_j)$ represents the corresponding probability of receiving a NACK by PU.

For the PU to fall in outage, its transmission rate $ \bar{r}_p $ must be greater than the channel instantaneous capacity $ C_p(i) $.
\begin{equation}
\Pr({\rm outage}|P_{\rm sec}=P_j)=\Pr(\bar{r}_p\emph{} > C_p(i)).
\end{equation}
where $C_p(i)$ for PU power $P$ and noise variance $N_0$ is given by
\begin{equation}
C_p(i)=B\log_2\left (1+ \frac{P|h_{pp}|^2}{N_0+P_{j}|h_{sp}|^2}\right) ,\quad j=0,1,2
\end{equation}
$ h_{pp} $ denotes the channel gain coefficient of the primary link and $ h_{sp} $ is the channel gain coefficient between the SU Tx and the PU Rx.

Let $ r_p = 2^{\bar{r_p}/B}-1$, with simple manipulation we get,
\begin{equation}
 \Pr({\rm outage}|P_{\rm sec}=P_j)=\Pr\left \{ r_p > \frac{P|h_{pp}|^2}{N_0+P_{j}|h_{sp}|^2} \right \}
\end{equation}
Assuming a Rayleigh fading distribution, the probability of outage or probability of NACK (interchangeably) has been derived in Appendix and expressed as:
\begin{equation}
\Pr({\rm outage}|P_{\rm sec}=P_j)= 1- e^{-\delta_{pp}r_pN_0/P}\left ( \frac{1}{1+\frac{\delta_{pp}P_jr_p}{P\delta_{sp}}} \right ).
\label{eqn:outage_prob}
\end{equation}
where $\delta_{pp}$ and $\delta_{sp}$ are parameters for exponential random variables representing the fading distributions for the respective primary and secondary links.

Due to error $\epsilon$ in accessing feedback, the probability of accessing a NACK by SU, $\Pr({\rm NACK}_j^{\rm a})$ is given by
\begin{equation}
\Pr({\rm NACK}_j^{\rm a})= (1-\epsilon) \Pr({\rm outage}|P_{\rm sec}=P_j)\quad j=0,1,2
\end{equation}
The steady state probabilities of the system Markov chain will be affected by imperfect feedback access as we explain next.
By characterizing the complete matrix of transition probabilities $ \mathbf{R} $, we get the steady state probabilities of the system Markov chain to calculate the PU probability of success.

\begin{figure}
\centering
\includegraphics[width=2.5in]{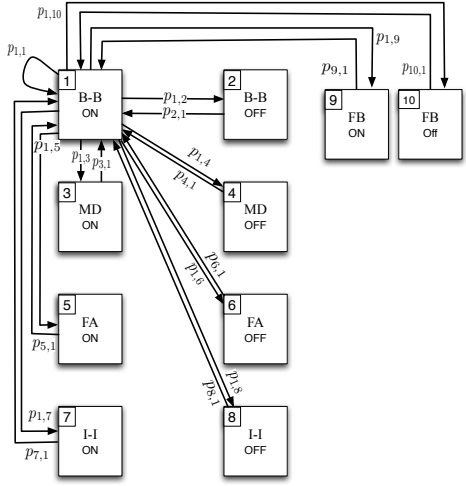}
\caption{The Markov chain model for both DPL and TPL schemes. However, transition probabilities vary for both schemes.} \label{MarcovCHFB}
\vspace{-0.3cm}
\end{figure}

The TPL scheme can be analyzed with the help of 10-state Markov chain as shown in Fig. \ref{MarcovCHFB}. These states represent channel dynamics of the SU. Each state represents the sensing process outcome, secondary link outage, SU access to feedback and PU activity. In particular, states from 1 to 8 model the SU when the PU accesses the medium with prior probability $ \rho $ as mentioned in section \ref{sysmod}. To model the sensing process outcome, we need 4 states, namely, (B-B, MD, FA and I-I). Similarly, we need 4 additional states to capture the ON/OFF channel states. On the other hand, states 9 and 10 represent the SU behaviour when it accesses a NACK from the PU receiver.

In order to fully characterize our Markov chain model, the transition probabilities of our model are characterized as follows:
\begin{eqnarray}
p_{1,1} &=&  \rho P_d \Pr (  r_1<C_1(i+TB)|r_1<C_1(i) )\nonumber\\
&& (1-(1-\epsilon)\Pr(\rm NACK_1))\\ &
=&\rho P_d \Pr ( z(i+TB)>\alpha _1 |z(i)> \alpha _1  )\nonumber\\
&& (1-(1-\epsilon)\Pr(\rm NACK_1))
\end{eqnarray}
where $ \alpha _1 = \frac {2^{\frac{r_1}{B}}}{\rm SNR_1}$. $ \Pr (  r_1<C_1(i+TB)|r_1<C_1(i) ) $ represents the probability that the secondary channel is ON (SU not in outage) and  $ (1-\Pr(\rm NACK_1)) $ is the probability of success when SU transmits with power $P_1$. In a block fading channel model, the fading changes independently from one block to another. Hence, $p_{1,1}$ can be further simplified as \cite{akin2010effective}
\begin{eqnarray}
p_{1,1}&=&  \rho P_d \Pr ( z(i+TB)> \alpha _1 ) (1-(1-\epsilon)\Pr(\rm NACK_1)) \nonumber\\ &=& \rho P_d P ( z> \alpha _1 ) (1-(1-\epsilon)\Pr(\rm NACK_1)).
\end{eqnarray}
Similarly,
\begin{eqnarray}
p_{i,1} =  \begin{cases}
  p_1=\rho P_d \Pr ( z > \alpha _1 ),  & i=5, 6, ..., 10 \\
   p_1(1-(1-\epsilon)\Pr(\rm NACK_1)),& i=1,2 \\
    p_1(1-(1-\epsilon)\Pr(\rm NACK_2)),&i=3,4
       \end{cases} .
\end{eqnarray}
\begin{eqnarray}
p_{i,2} = \begin{cases}
  p_2=\rho P_d \Pr ( z < \alpha _1 ),  &i=5, 6, ..., 10 \\
   p_2(1-(1-\epsilon)\Pr(\rm NACK_1)),&i=1,2 \\
    p_2(1-(1-\epsilon)\Pr(\rm NACK_2)),&i=3,4
       \end{cases}
\end{eqnarray}
Similarly,
\begin{equation}
p_{i,3}= p_3 = \rho (1-P_d) \Pr ( z > \alpha _2 ),   \;\;\;\;   \;\;\;\; i=5,6,..,10.
\end{equation}
where $ \alpha _2 = \frac {2^{\frac{r_2}{B}}}{\rm SNR_2}$, $ \alpha _3 = \frac {2^{\frac{r_1}{B}}}{\rm SNR_3}$, $ \alpha _4 = \frac {2^{\frac{r_2}{B}}}{\rm SNR_4}$ and $ \alpha _5 = \frac {2^{\frac{r_0}{B}}}{\rm SNR_5}$.

As in states 1 and 2, other transition probabilities for states $4,5,\cdots,8$ can be characterized in the same way\footnote{Transition probabilities for states $4,5,\dots,8$ are omitted due to space limitations.}. However, for states 9, 10 the transition probabilities are different since the probability that the system enters these states as a function of the PU outage probability.
\begin{equation}
p_{i,9} = \begin{cases} \label{eqn_9}
 (1-\epsilon) \Pr(\rm NACK_1)  \Pr ( z > \alpha _5 ),  &\mbox{$i=1,2$} \\
    (1-\epsilon)\Pr(\rm NACK_2)  \Pr ( z > \alpha _5 ),&\mbox{$i=3,4$}  \\
0, &\mbox{\rm otherwise}
       \end{cases} .
\end{equation}
\begin{equation}
p_{i,10} =  \begin{cases} \label{eqn_10}
   (1-\epsilon)\Pr(\rm NACK_1)  \Pr ( z < \alpha _5 ),  &\mbox{$i=1,2$} \\
    (1-\epsilon)\Pr(\rm NACK_2)  \Pr ( z < \alpha _5 ),&\mbox{$i=3,4$}  \\
0, &\mbox{\rm otherwise}
       \end{cases} .
\end{equation}
We assume that the system will never stay in states 9 or 10 for 2 successive time frames due to buffer limitations. It is also clear that no transitions are permitted from states 5 up to 8 to states 9 and 10 as the PU can never receive a NACK message while being silent.

Accordingly, we have completely specified the transition probability matrix $\mathbf{R}_{10 \times 10}$. The moment generating function corresponding to each state depends on the effective rate of each state \cite{mypaper}. Hence,
 \begin{equation}
\begin{split}
\mathbf{\Phi}(-\theta) &= diag\lbrace e^{-(T-N) \theta r_1}, 1, e^{-(T-N) \theta r_2} \\& , 1, e^{-(T-N) \theta r_1}, 1, e^{-(T-N) \theta r_2}, 1, e^{-T \theta r_0}, 1  \rbrace.
\end{split}
\end{equation}

\subsection{Quantifying Primary User Performance}
\label{sect:success rate}
In order to study the cognitive network fundamental tradeoff, we quantify the expectation of receiving a NACK for TPL scheme.

The average outage (NACK) probability can be expressed by Bayes theorem as:
\begin{equation}\label{basic_eqn}
\Pr({\rm NACK}) = \sum _{j=0}^2  \Pr({\rm outage}|P_{\rm sec}=P_j)\times  \Pr(P_{\rm sec}=P_j).
\end{equation}
As a result of FSMC model explained in previous section,
\begin{eqnarray}
\Pr( {P_{\rm sec}= P_0}) &=& \sum _{i=9,10} \beta_i\\ \label{eqn_outage}
\Pr( {P_{\rm sec}= P_1}) &=& \sum _{i=1,2} \beta_i\\
\Pr({ P_{\rm sec}= P_2}) &=& \sum _{i=3,4} \beta_i
\end{eqnarray}
where,
\begin{equation}
\beta_i = \frac{\pi_i}{\sum _{i=1,2,3,4,9,10} \pi_i}.
\end{equation}
$ \pi_i $ is the steady state probability of each state in FSMC. Although SU also sends with power levels $ P_1 $ and $ P_2 $  in states 5,6 and 7,8, respectively, the PU is idle in these states (i.e, cannot be in outage). Since, $\sum_{j=0}^2\Pr({P_{sec}=P_j}) $ should sum to one, $ \beta_i $ is the normalized (by the states where PU transmits data) value of the steady state $ \pi_i $.

\section{Numerical Results}
\label{sect:results}
\begin{table}
\caption{Parameters' numerical values}
\label{parameters}
    \begin{tabular}{|c c c |}
    \hline
        $r_1=1000 $ bps  &  $r_2=2000$ bps    & $r_p=100000 $ bps        \\
        $P_1/B=0.25$ W/Hz &  $P_2/B=1$  W/Hz&  $P_{pu}/B=100$ W/Hz       \\
        $B=100$ KHz  &  $\rho = 0.1$       & $ \lambda=1.85 $              \\
        $T=0.01$ sec & $N=0.003$ sec & $N_0=\sigma_n^2=1$   \\
        $\delta_{pp}=0.1$  & $\delta_{sp}=0.1$    &$\epsilon=0.3$  \\
        \hline
    \end{tabular}
    \vspace{-0.3cm}
\end{table}
\begin{figure}
\centering
    {\includegraphics[width=3.2in]{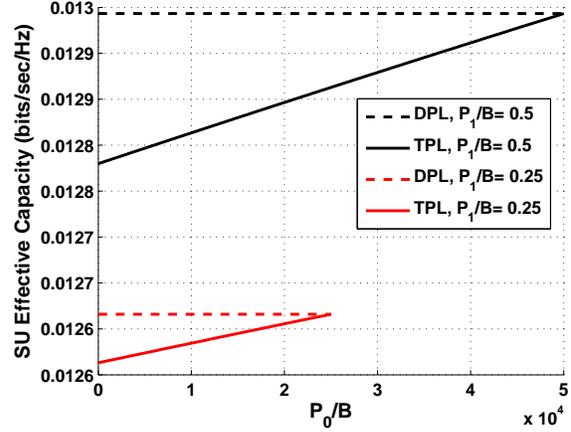}
     \vspace{-0.4cm}
  \caption{Secondary user EC for error free feedback access case for both DPL and TPL schemes.}
  \vspace{-0.4cm}
 \label{fig:EC_region}}
 \end{figure}


In this section, we evaluate the tradeoff between secondary link EC and PU success rate for both DPL and TPL schemes. For the numerical results, we use the parameter values described in Table. \ref{parameters}. Delay exponent $\theta$ equals 0.01 for all numerical results.

In Fig. \ref{fig:EC_region}, we plot the EC of the SU for both DPL and TPL schemes for a fixed value of $P_1$. $P_0$ for TPL is increased unless it reaches $P_1$ which is the same for both DPL and TPL. Please note that DPL has no third power level $P_0$. Therefore, its performance is independent of $P_0$ and works as a bench mark in our numerical evaluations. Introduction of third power level in TPL while retransmitting a PU packet reduces effective capacity of the SU for TPL as compared to DPL. When $ P_0 $ varies from zero to $ P_1 $, the EC of the SU increases for TPL and becomes equal to EC of DPL when $P_0=P_1$. Obviously, $P_0=0$ provides the smallest EC as SU does not transmit at all during retransmission phase of PU and causes no interference to PU. When $P_1$ is large as compared to $P_0$, the effect of introducing $P_0$ is more pronounced.
\begin{figure}
\centering
    {\includegraphics[width=3.2in]{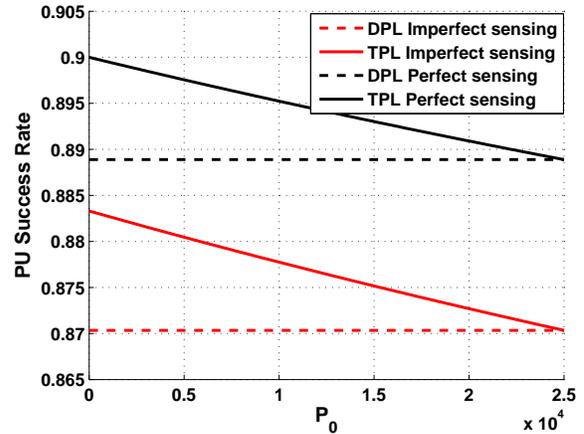}
     \vspace{-0.3cm}
  \caption{PU Success rate for error free feedback access case for both DPL and TPL schemes.}
 \label{fig:success_rate_perfect}}
 \vspace{-0.4cm}
 \end{figure}

In Fig.~\ref{fig:success_rate_perfect}, we observe the corresponding effects on PU packet success rate for a fixed $P_1$. We consider both perfect and imperfect sensing cases for both DPL and TPL. Corresponding to decrease of EC for SU with increasing $P_0$, TPL scheme increases PU's success rate as compared to DPL. Increasing $P_0$ increases interference for PU and decreases his packet success rate. The success rate for the perfect sensing case is higher than the imperfect sensing case because the perfect sensing allows SU to transmit with a lower power level if the PU is transmitting.

\begin{figure}
{\includegraphics[width=3.2in]{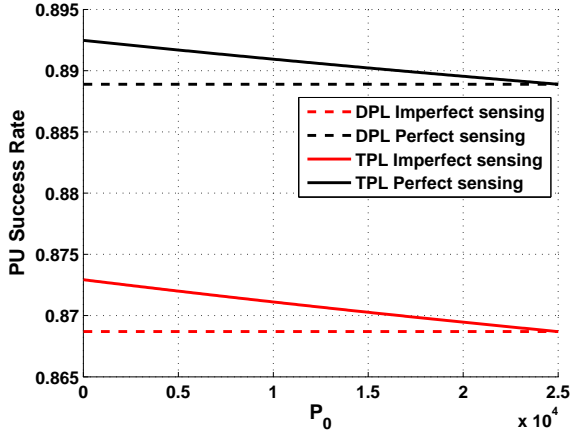}
\vspace{-0.3cm}
\caption{PU Success rate for the erroneous feedback access case for the both DPL and TPL schemes.}
  \label{fig:success_rate_imperfect}}
  \vspace{-0.4cm}
\end{figure}
\begin{figure}
\centering
    {\includegraphics[width=3.2in]{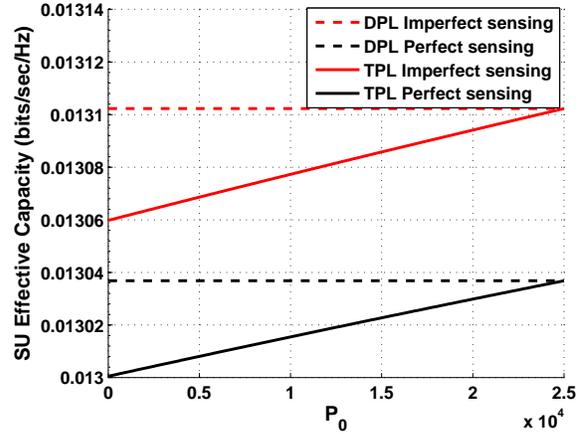}
    \vspace{-0.3cm}
  \caption{Secondary user EC for the erroneous feedback access case for the both DPL and TPL schemes. }
 \label{fig:EC_region_imperfect}}
\vspace{-0.4cm}
 \end{figure}

For the same problem settings and erroneous SU access to PU feedback, Fig.~\ref{fig:success_rate_imperfect} shows success rate of PU. For a fixed error in feedback access $\epsilon$, success rate of PU decreases as error in feedback access causes SU to transmit with higher power level (erroneously) during the PU retransmission phase. As we characterized in Section \ref{sect:Schemes}, TPL scheme is affected severely as compared to DPL by erroneous access of PU feedback. By comparing the PU success rate for erroneous feedback access with perfect feedback access case, we verify numerically that PU success rate (and SU EC) is not affected by erroneous feedback access for DPL if sensing is perfect as characterized in Section \ref{sect:Schemes}. Similar conclusions are drawn for EC of SU in Fig.~\ref{fig:EC_region_imperfect} for the erroneous feedback access case where EC increases as compared to the perfect feedback access case for all values of $P_0$ for both of the schemes.


\section{Conclusions}\label{sect:conclusions}
In this paper, we analyze the performance of a cognitive system consisting of one secondary link and one primary network abstracted by a single user accessing a single channel with certain probability. We characterize the tradeoff between primary user success rate and secondary user effective capacity. We propose a three power level scheme to facilitate primary user transmission which outperforms two power level scheme presented earlier in literature in terms of primary user packet success rate. We analyze both the schemes in the scenario when secondary user accesses feedback for primary transmitter erroneously and validate the analysis numerically.

\section*{Appendix}
\label{sect:outage}
We first derive probability of success $ \Pr({\rm suc}|P_{\rm sec}=P_j)$ for an SU transmission power level $P_j$ and Rayleigh fading.
\begin{eqnarray}
\Pr({\rm suc}|P_{sec}=P_j)&=& \! \Pr\left \{ r_ p <\! \frac{P |h_{pp}|^2}{N_0\!+\! P_j|h_{ sp}|^2} \right \}\\
&=&\Pr\left \{\!|h_{pp}|^2  >\! \frac{ r_{ p}\! }{P} \left ( N_0\!+\! P_j|h_{sp}|^2 \right )\right \}\nonumber\\
&=&\Pr\left \{\!\chi_{pp}  >\! \frac{ r_{p}\! }{P} \left ( N_0\!+\! P_j.\chi_{sp} \right )\right \}\nonumber
\end{eqnarray}
Both random variables $\chi_{pp} $ and $\chi_{sp}$ are exponentially distributed with parameters $\delta_{pp}$ and $\delta_{sp}$, respectively.

Thus, $\Pr({\rm suc}|P_{sec}=P_j)$ is derived as
\begin{eqnarray}
 &&\Pr({\rm suc}|P_{\rm sec}=P_j)=\\
&&\int_{x_{sp}=0}^{\infty}\Pr\left ( \chi_{pp}  >\! \frac{ r_{p}\! }{P} \left ( N_0\!+\! P_jx_{sp} \right ) \right ) \times \delta_{sp}
e^{-\delta_{sp}x_{sp}} dx_{sp}\nonumber\\
&=&\int_{x_{sp}=0}^{\infty} e^{-\delta_{pp}\left ( \frac{r_pN_0}{P}+\frac{P_jr_p}{P}x_{sp} \right )}\times \delta_{sp} e^{-\delta_{sp}x_{sp}} dx_{sp}\nonumber\\
&=&\int_{x_{sp}=0}^{\infty} e^{-\delta_{pp}\frac{r_pN_0}{P}} \times \delta_{sp} e^{-\left (\frac{\delta_{pp}P_jr_p}{P}+\delta_{sp}  \right )x_{sp}} dx_{sp}
\end{eqnarray}
Solving the above integration, $\Pr({\rm suc}|P_{sec}=P_j)$ is given by
\begin{equation}
\Pr({\rm suc}|P_{\rm sec}=P_j)= e^{-\delta_{pp}r_pN_0/P}\left ( \frac{1}{1+\frac{\delta_{pp}P_jr_p}{P\delta_{sp}}} \right ).
\label{eqn:sucess_prob}
\end{equation}
Using (\ref{eqn:sucess_prob}) in (\ref{eqn:pr_nack}) yields (\ref{eqn:outage_prob}).
\section*{Acknowledgement}
This paper was made possible by a NPRP grant 4-1034-2-385 from the Qatar National Research Fund (a member of The
Qatar Foundation). The statements made herein are solely the responsibility of the authors.
\bibliographystyle{IEEEtran}
\bibliography{bibliography}

\end{document}